\documentclass[fleqn,twoside,twocolumn,nofootinbib,showkeys]{revtex4} 
\usepackage[sec,nocpr]{ujp} 

\begin{document}
\title[Electronic Structure and Properties]
{ELECTRONIC STRUCTURE AND PROPERTIES\\ OF NOVEL LAYERED SUPERCONDUCTORS}
\author{G.E.~Grechnev}
\affiliation{B. Verkin Institute for Low Temperature Physics and
Engineering, Nat. Acad. of Sci. of Ukraine}
\address{47, Lenin Ave., Kharkov 61103, Ukraine}
\email{grechnev@ilt.kharkov.ua}
\author{A.V.~Logosha}
\affiliation{B. Verkin Institute for Low Temperature Physics and
Engineering, Nat. Acad. of Sci. of Ukraine}%
\address{47, Lenin Ave., Kharkov 61103, Ukraine}%
\author{A.A.~Lyogenkaya}
\affiliation{B. Verkin Institute for Low Temperature Physics and
Engineering, Nat. Acad. of Sci. of Ukraine}%
\address{47, Lenin Ave., Kharkov 61103, Ukraine}%
\author{A.G.~Grechnev}%
\affiliation{B. Verkin Institute for Low Temperature Physics and
Engineering, Nat. Acad. of Sci. of Ukraine}%
\address{47, Lenin Ave., Kharkov 61103, Ukraine}%
\author{A.V.~Fedorchenko}
\affiliation{B. Verkin Institute for Low Temperature Physics and
Engineering, Nat. Acad. of Sci. of Ukraine}
\address{47, Lenin Ave., Kharkov 61103, Ukraine}
 \udk{538.945} \pacs{74.20.Pq, 74.62.Fj,\\[-3pt] 74.70.Dd, 74.70.Xa,\\[-3pt] 75.10.Lp} \razd{\secvii}

\autorcol{G.E.\hspace*{0.7mm}Grechnev, A.V\hspace*{0.7mm}Logosha,
A.A\hspace*{0.7mm}Lyogenkaya et al.}

\setcounter{page}{284}%

\begin{abstract}
The electronic energy structures and magnetic properties of layered
superconductors $R$Ni$_2$B$_2$C, $R$Fe$_4$Al$_8,$ and FeSe are
systematically studied, by using the density functional theory
(DFT). The calculations allowed us to reveal a number of features of
the electronic structure, which can cause the manifestation of
peculiar structural, magnetic and superconducting properties of
these systems. It is demonstrated that the Fermi energy $E_{\rm F}$
is located close to the pronounced peaks of the electronic density
of states (DOS).\,\,The main contribution to DOS at the Fermi level
arises from $3d$-electrons.\,\,The calculations of the
pressure-dependent electronic structure and the magnetic
susceptibility in the normal state indicate that the novel
superconductors are very close to a magnetic instability with
dominating spin paramagnetism.\,\,It is shown that experimental data
on the pressure dependence of the superconducting transition
temperature in FeSe correlate qualitatively with the calculated
behavior of DOS at $E_{\rm F}$ as a function of the pressure.
\end{abstract}
 \maketitle

\section{Introduction}
The discovery of the superconductivity in transition metal
borocarbides with the general formula $R$Ni$_2$B$_2$C ($R =$Y, Ho,
Er, Tm or Lu) and its coexistence with magnetism stimulated a
considerable scientific interest in these systems \cite
{muller08}.\,\,Later, the superconductivity was also found in
magnetic compounds YFe$_4$Al$_8$, LuFe$_4$Al$_8,$ and ScFe$_4$Al$_8$
at temperatures lower than 6 K \cite{dmitriev03}.\,\,In 2008, the
new class of iron-based layered superconductors was
discovered.\,\,One of the representatives of this class is FeSe
compound distinguished by the simplest crystal structure among
iron-based superconductors and by the extremely large effect of a
pressure on the superconducting transition temperature \cite
{mizuguchi10,wen11,bendele12}.

The characteristic feature of these layered compounds of $3d$-metals
is the well-established coexistence of magnetism and superconductivity.
The relative structural simplicity of $R$Ni$_2$B$_2$C, $R$Fe$_4$Al$_8,$
and FeSe is favourable for studying the effects of chemical substitution,
high pressure, and uniaxial deformations on their physical properties.
Such studies can reveal a mechanism of superconductivity in these systems,
which contain magnetic $3d$-metals.

The clarification of microscopic mechanisms, which determine
electric and magnetic properties of metallic systems, assumes
detailed experimental and theoretical studies of the electronic
structure of a conduction band.\,\,A number of electronic structure
calculations were carried out for nickel borocarbides \cite
{divis00,shorikov06} and superconducting FeSe
\cite{subedi08,pickett08} in recent years. However, the data on the
electronic energy structure of these systems are still incomplete
and inconsistent.

In very recent studies of magnetic superconductors with the angular
resolved photoemission spectroscopy method (ARPES) \cite
{kordyuk12}, the presence of Van Hove singularities was established
in the electronic structures in a small vicinity of the Fermi energy
$E_{\rm F}$. Moreover, the technological progress in growing
single-crystal samples provided an opportunity to study the fine
structure of the electronic energy bands and the Fermi surface of
nickel borocarbides by means of the de Haas--van Alphen effect
\cite{goll96,bergk09} and ARPES \cite{baba10} and FeSe compound
(ARPES, \cite{borisenko13}). Thus, the detailed {\it ab initio}
calculations of the electronic structure are necessary for the
analysis of the spectral characteristics of these systems.

Here, we have calculated the electronic band structures and a number of thermodynamic
characteristics of $R$Ni$_2$B$_2$C, $R$Fe$_4$Al$_8,$ and FeSe compounds within the density
functional theory (DFT) methods.
The dependences of these characteristics on the volume and structural parameters
were addressed to shed light on the corresponding high-pressure effects.

\section{Details and Results\\ of Electronic Structure Calculations}
The calculations of electronic structures were carried out by using the modified relativistic
LMTO method with a full potential (FP-LMTO, RSPt implementation \cite{grechnev09,wills10})
and the linearized augmented plane waves method with a full potential
(FP-LAPW, Elk implementation \cite{elk}).
The exchange and correlation potentials were treated within the local density approximation (LDA
\cite{barth72}) and the generalized gradient approximation (GGA \cite{pbe96})
of DFT.
For the employed full potential FP-LMTO and FP-LAPW methods, any
restrictions were not imposed on the charge densities or potentials of the studied systems,
which is especially important for anisotropic layered structures of the investigated
magnetic superconductors.

Electronic structure calculations for compounds were carried out for
the sets of crystal lattice parameters close to experimental ones.
Variations of the parameters of crystal lattices allow one to imitate
the influence of an external pressure. In such a way, the volume
dependence of the total energy of the electronic subsystem,
$E(V)$, was calculated for the studied compounds. The theoretical
values of equilibrium lattice parameters and bulk moduli $B$ were
determined from the calculated $E(V)$ dependences, by using the known
Birch--Murnaghan equation of state (see Refs.
\cite{grechnev09,wills10}).

\begin{figure}%
\vskip1mm
\includegraphics[width=6cm]{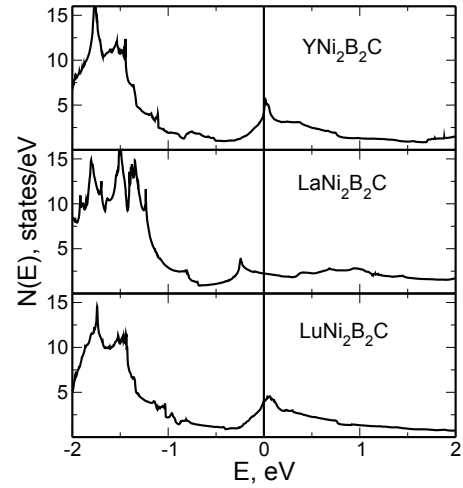}
\vskip-3mm\caption{Density of electronic states $N(E)$ of
YNi$_2$B$_2$C, LaNi$_2$B$_2$C, and LuNi$_2$B$_2$C compounds. The
Fermi level ($E=$ $=0$) is marked by the vertical line  }
\end{figure}

To evaluate the paramagnetic susceptibility of compounds,
the FP-LMTO calculations of field-induced spin and orbital
(Van Vleck) magnetic moments were carried out with the approach
described in Ref. \cite{grechnev09} within the local
spin density approximation (LSDA) of DFT.
The relativistic effects, including spin-orbit coupling, were incorporated,
and the effect of an external magnetic field $\mathbf{B}$ was taken into account
self-consistently by means of the Zeeman term,
\begin{equation}
\label{zeeman} {\cal H}_{Z}=\mu_{\rm B} \mathbf{B}
(2\hat{\bf{s}}+\hat{\bf{l}}).
\end{equation}
Here, $\mu_{\rm B}$ is the Bohr magneton, and $\hat{\bf{s}}$ and $\hat{\bf{l}}$
are the spin and orbital angular momentum operators, respectively.
The field-induced spin and orbital magnetic moments provide the related contributions
to the magnetic susceptibility, $\chi_{\rm spin}$ and $\chi_{\rm orb}$.

\subsection{\boldmath$R$Ni$_2$B$_2$C}

The crystal structure of nickel borocarbides (like YNi$_2$B$_2$C, space group $I4/mmm$)
is a body-centered tetragonal structure with alternating triple layers of B-Ni-B
and Y-C planes.
In this work, we carried out the DFT calculations of the band structures,
densities of electronic states (DOS), and some thermodynamic characteristics
of YNi$_2$B$_2$C, LaNi$_2$B$_2$C, and LuNi$_2$B$_2$C compounds.
These compounds contain non-magnetic trivalent transition metals Y, La, and Lu,
whose external electronic shells are similar to those of rare-earth elements $R$.
For each compound, the calculations of the electronic structure were carried out
for a number of lattice parameters, close to experimental ones
(listed in Refs. \cite{muller08,lynn97}).
At the same time, the $c/a$ ratio was fixed and corresponded to the experimental
value for each $R$Ni$_2$B$_2$C compound.

\begin{figure}%
\vskip1mm
\includegraphics[width=5cm]{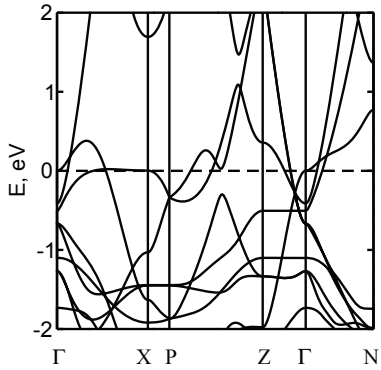}
\vskip-3mm\caption{Band structure of YNi$_2$B$_2$C along high
symmetry lines of the Brillouin zone. The Fermi level ($E=0$) is
marked by the horizontal line  }
\end{figure}

The calculated densities of electronic states $N(E)$ of these nickel
borocarbides are similar, but differ in details and the positions of
Fermi levels $E_{\rm F}$ (see Fig.~1). In Fig.~1, one can see the
sharp peaks in $N(E)$ of YNi$_2$B$_2$C and LuNi$_2$B$_2$C
superconductors in a close proximity of the Fermi level, whereas the
related peak in the non-superconducting LaNi$_2$B$_2$C compound is
situated much deeper below $E_{\rm F}$.

\begin{table*}[!]
\vskip4mm \noindent\caption{Thermodynamic properties of
\boldmath$R$Ni$_2$B$_2$C borocarbides ($R$=Y, La, Lu).\\ $V$ -- unit
cell volume, $N(E_{\rm F})$ -- DOS at the Fermi level, $\gamma$ --
electronic specific heat (Ref. \cite{michor95}),\\ $\Theta_{\rm D}$
-- Debye temperature (Ref. \cite{michor95}), $\lambda$ --
electron-phonon mass enhancement,\\ $T_{\rm c}$ --  superconducting
transition temperature (from \cite{michor95,budko00})
}\vskip3mm\tabcolsep6.0pt

\noindent{\footnotesize\begin{tabular}{|l|c|c|c|c|c|c|c|}
  \hline
  \parbox[c][11mm][c]{20mm}{Combination} &
  \parbox[c][11mm][c]{15mm}{$V$, {\AA}$^3$ } &
  \parbox[c][11mm][c]{20mm}{$N(E_{\rm F })$, states/eV$\cdot$cell } &
  \parbox[c][11mm][c]{20mm}{$\gamma (\exp )$, mJ/mol$\cdot$K$^2$ } &
  \parbox[c][11mm][c]{15mm}{$\Theta _{\rm D}$, K } &
  \parbox[c][11mm][c]{10mm}{$\lambda$ } &
  \parbox[c][11mm][c]{15mm}{$T_c (\exp )$, K } &
  \parbox[c][11mm][c]{18mm}{$T_c $(theor), K} \\
  \hline \rule{0pt}{5mm}~~~YNi$_2$B$_2$C & 65.9 & 4.30 & 18.2 & 490 &~0.8~ & 15.6 & 15.3 \\
~~~LaNi$_2$B$_2$C & 70.7 & 2.24 & 8.4 & 495 &~0.6~ & --- & ~\,6.5 \\
~~~LuNi$_2$B$_2$C & 63.8 & 4.07 & 19.5 & 360 &~1.0~ & 16.6 & 17.8 \\[2mm]
  \hline
\end{tabular}}
\end{table*}

The calculated band structure of YNi$_2$B$_2$C compound is presented
in Fig.~2, where one can see the presence of quasi-degenerate states
near the Fermi level in a vicinity of the symmetry point $\Gamma$
and at PZ line and the almost dispersion-free branch of $E(k)$ in
the direction $\Gamma$X. The position of this branch corresponds to
the sharp peak of DOS in a vicinity of $E_{\rm F}$ (Van Hove
singularity in Fig.~1). The main contribution to $N(E_{\rm F})$
comes from the $d$-states of nickel layers.

The reliability of the present calculations of the electronic structure
of nickel borocarbides is confirmed by a good description of the
low-temperature quantum oscillation of magnetization -- the de
Haas--van Alphen effect (DHVA)~-- for YNi$_2$B$_2$C and
LuNi$_2$B$_2$C compounds \cite{goll96,bergk09}. In particular, by
comparison with the DHVA experimental data for YNi$_2$B$_2$C and
LuNi$_2$B$_2$C, it is established that the low-frequency branches of
DHVA oscillations found in \cite{goll96,bergk09},
$F_{\alpha}\simeq$~500 T, correspond to the Fermi surface
sections in a vicinity of the symmetry point $\Gamma$ (see $E(k)$ dependences
in Fig.~2). It should be noted that, for a small shift of the Fermi level
(less than 0.1~eV), which corresponds to the accuracy of the {\it ab initio} calculations of the Fermi energy position, the calculated
sections of YNi$_2$B$_2$C and LuNi$_2$B$_2$C Fermi surfaces appeared
to be in agreement with $F_{\alpha}$ data from Refs.
\cite{goll96,bergk09} within the experimental error. The ratio of
experimental cyclotron masses to corresponding calculated ones,
$m^c_{\rm exp}/m^c_ {\rm theor}=(1 +\lambda)$, for the Fermi surface
section $F_{\alpha}$ of YNi$_2$B$_2$C is within the limits of
$1.5\div1.7$ for various directions of the magnetic field. This
provides the reasonable value of the corresponding constant of
many-body mass enhancement, $\lambda_{\alpha}\simeq$~0.6.

The calculated values of DOS at the Fermi level $N(E_{\rm F})$ for
the nickel borocarbides are presented in Table~1 and can be compared
with the available experimental data on the electronic specific heat
coefficients $\gamma_{\rm exp}$ \cite {michor95}. The differences
between $\gamma_{\rm theor}$ and $\gamma_{\rm exp}$ are usually
attributed to the renormalization of one-electron effective masses due
to the electron-phonon interaction,
\begin{equation}
\label{gamma_lambda}
\gamma_{\rm exp}=(1 +\lambda) \gamma_{\rm theor},
\end{equation}
which gives a possibility to determine the corresponding
renormalization parameter $\lambda$ (see Table~1).

By using the theoretical and experimental data from Table 1, one can
estimate the superconducting transition temperatures of the
investigated nickel borocarbides with the use of the Macmillan formula
\cite{mcmillan68}:
\begin{equation}
\label{mcmillan}
T_{\rm c} = \frac{\Theta_{\rm D}}{1.45}\exp \biggl[-\frac{1.04(1+\lambda)}
{\lambda-\mu^{\ast}(1+0.62\lambda) }\biggr],
\end{equation}
where $\Theta_{\rm D}$ -- Debye's temperature, $\lambda$ --
electron-phonon interaction constant, and $\mu^{\ast}$ --
Morel--Anderson's Coulomb pseudopotential. The value of $\mu^{\ast}$
is taken to be 0.13, as accepted for transitional metals \cite
{mcmillan68}. Thus, using the experimental values of Debye
temperatures $\Theta_{\rm D}$ and estimated $\lambda,$ we obtained
$T_c$ values, which are in a good agreement with experimental data
for YNi$_2$B$_2$C and LuNi$_2$B$_2$C (see Table~1). The difference of
the estimated $T_c$ with the experimental value for LaNi$_2$B$_2$C can be caused
either by errors in the evaluations of $\gamma_{\rm exp}$ and $N(E_{\rm F})$ or due to a large spin-fluctuation contribution
$\lambda_{\rm sf}$ to the renormalization parameter $\lambda$ in
(\ref{gamma_lambda}):
\begin{equation}
\label{sf}
\lambda = \lambda_{\rm el-ph}+\lambda_{\rm sf},
\end{equation}
which can explain a smaller contribution of $\lambda_{\rm el-ph}$.
Nevertheless, the results in Table~1 clearly testify in favor of the
BCS-like electron-phonon mechanism of superconductivity in the
nickel borocarbides, \mbox{having $\lambda \approx1$.}

The FP-LMTO calculations of the field-induced spin and orbital (Van
Vleck) magnetic moments were carried out for YNi$_2$B$_2$C,
LaNi$_2$B$_2$C, and LuNi$_2$B$_2$C compounds with inclusion of the
Zeeman operator (\ref{zeeman}) in the external magnetic field
\mbox{$\textbf{B}=10$~T}. For the tetragonal crystal structure, the
corresponding contributions to the magnetic susceptibility,
$\chi_{\rm spin}$ and $\chi_{\rm orb}$, were calculated for the
external field directed along the $c$ axis. The values of magnetic
susceptibility ($\chi_{\rm theor}=\chi_{\rm spin}+\chi_{\rm orb}$)
calculated for YNi$_2$B$_2$C, LaNi$_2$B$_2$C, and LuNi$_2$B$_2$C are
listed in Table~2 in comparison with the available experimental
data.

In the general form, the total magnetic susceptibility can be decomposed
in the following terms (see Ref. \cite{grechnev09}):
\begin{equation}
\label{chitot} \chi_{\rm tot} =\chi_{\rm spin} +\chi_{\rm orb}
+\chi_{\rm dia} +\chi_{\rm L} ,
\end{equation}
which present, respectively, Pauli's spin susceptibility ($\chi_{\rm
spin}$), Van Vleck orbital paramagnetism ($\chi_{\rm orb}$),
Langevin diamagnetism of closed ion shells ($\chi_{\rm dia}$), and
the orbital diamagnetism of conduction electrons ($\chi_{\rm
L}$).\,\,It can be seen in Table~2 that the spin and orbital Van
Vleck susceptibilities are the principal terms in
Eq.~(\ref{chitot}).\,\,It is important that $\chi_{\rm orb}$ gives a
substantial contribution to the full paramagnetic susceptibility of
borocarbides.\,\,The theoretical calculation of the Landau
diamagnetism $\chi_{\rm L}$ for the multiband dispersion law $E(k)$
represents a very difficult task \cite {grechnev09}.\,\,However, a
good fit of the experimental susceptibilities of YNi$_2$B$_2$C and
LuNi$_2$B$_2$C with the calculated paramagnetic contributions to
$\chi$ (see Table~2) indicates that, in these superconducting
systems, the diamagnetic contributions in (\ref{chitot}) are
insignificant.

\begin{table}[t]

\vskip4mm \noindent\caption{Magnetic susceptibility\\ of
\boldmath$R$Ni$_2$B$_2$C ($R=Y$, La, Lu)  }\vskip3mm\tabcolsep10.3pt

\noindent{\footnotesize\begin{tabular}{|l|c|c|c|c|}
  \hline
   \multicolumn{1}{|c}
{\raisebox{-3mm}[0cm][0cm]{Compound}} &
\multicolumn{1}{|c}{\rule{0pt}{5mm}$\chi_{\rm spin}$ }&
\multicolumn{1}{|c}{$\chi_{\rm orb}$}&
\multicolumn{1}{|c}{$\chi_{\rm theor}$}&
\multicolumn{1}{|c|}{$\chi_{\exp}$}\\[2mm]
\cline{2-5}
 \multicolumn{1}{|c}{\rule{0pt}{5mm}}&
\multicolumn{4}{|c|}{$10^{-4}$ emu/mol} \\[2mm]
  \hline \rule{0pt}{5mm}YNi$_2$B$_2$C &  1.35 & 0.87 & 2.22 & 2.0 \cite{fisher95}\\
  LaNi$_2$B$_2$C & 0.96 & 0.72 & 1.68 & 1.0 \cite{fisher95} \\
  LuNi$_2$B$_2$C & 1.29 & 0.73 & 2.02 & 1.9 \cite{lai95} \\[2mm]
  \hline
\end{tabular}}
\end{table}

\subsection{\boldmath$R$Fe$_4$Al$_8$}
Compounds $R$Fe$_4$Al$_8$ possess the body-centered tet\-ragonal
crystal structure of the ThMn$_{12}$-type, which belongs to the
$I4/mmm$ space symmetry group \cite{suski04}. In case of
non-magnetic trivalent transition metals ($R=$~Sc, Y, Lu), the AFM
ordering of iron moments was established in $R$Fe$_4$Al$_8$ at low
temperatures, though there are very inconsistent experimental data
in the literature about a character of AFM ordering, ordering
temperatures, and magnetic moments (see
\cite{dmitriev03,suski04,gasche06,paixao01} and references therein).

Virtually, no band structure calculations have been carried out to
date for $R$Fe$_4$Al$_8$.\,\,In work \cite{gasche06}, the main
attention was paid to calculations of the iron magnetic moments in
the magnetically ordered phases of YFe$_4$Al$_8$.\,\,In the present
work, the DFT electronic calculations are carried out for
YFe$_4$Al$_8$, ScFe$_4$Al$_8,$ and LuFe$_4$Al$_8$ compounds in the
paramagnetic, ferromagnetic (FM), and antiferromagnetic (AFM)
phases.

\begin{figure}%
\vskip1mm
\includegraphics[width=6cm]{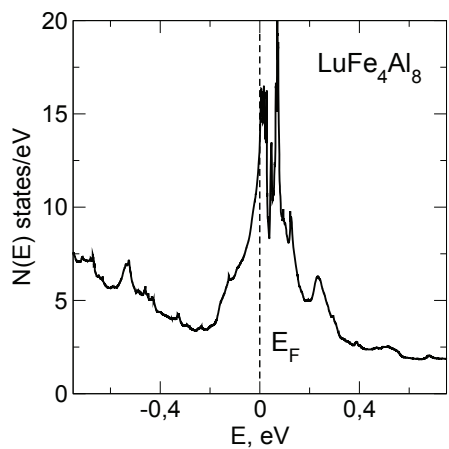}
\vskip-3mm\caption{Density of electronic states $N(E)$ of
LuFe$_4$Al$_8$ in the paramagnetic phase. The Fermi level ($E=0$) is
marked by the vertical line  }
\end{figure}

\begin{table}[b]
\noindent\caption{Densities of electronic states\\ at the Fermi
level and the magnetic moments of iron\\ atoms for
\boldmath$R$Fe$_4$Al$_8$ compounds ($R=$~Sc, Y, Lu)\\ in
the paramagnetic (PM), ferromagnetic (FM)\\ and antiferromagnetic (AFM)
phases }\vskip3mm\tabcolsep8.0pt

\noindent{\footnotesize\begin{tabular}{|l|c|c|c|c|c|}
  \hline
 \multicolumn{1}{|c}
{\raisebox{-5mm}[0cm][0cm]{Combination}} &
\multicolumn{3}{|c}{\rule{0pt}{5mm}$N(E_{\rm F})$}&
\multicolumn{2}{|c|}{$M$}\\%
\multicolumn{1}{|c}{}&\multicolumn{3}{|c}{states/(eV$\cdot$cell)}
&\multicolumn{2}{|c|}{$\mu_{\rm B}$}\\[2mm]
  \cline{2-6}
  \rule{0pt}{5mm}&PM&FM&AFM&FM&AFM\\[2mm] \hline%
  \rule{0pt}{5mm}~~ScFe$_4$Al$_8$ & 25.2 & 10.5 & 14.5 & 1.20 & 1.17 \\
 ~~YFe$_4$Al$_8$  & 26.5 & ~\,8.4 & 14.7 & 1.27 & 1.25 \\
~~LuFe$_4$Al$_8$ & 27.0 & 11.5 & 14.3 & 1.24 & 1.22 \\[2mm]
  \hline
\end{tabular}}
\end{table}

The calculated DOS of the paramagnetic phase of LuFe$_4$Al$_8$ is
presented in Fig.~3. The densities of states $N(E)$ for the
isoelectronic ScFe$_4$Al$_8$ and YFe$_4$Al$_8$ compounds are very
similar, and they differ in small details only. The calculated
densities of states at the Fermi level are listed in Table~3, and
the dominating contributions to $N(E_{\rm F})$ come from the $d$-states
of iron. As is seen from Fig.~3, the Fermi level for the PM phase of LuFe$_4$Al$_8$
is located at the steep slope of the density of
states, where $N(E)$ quickly grows with the energy in the immediate
proximity ($\sim$0.01 eV) to the sharp peak of DOS.

The peak of $N(E)$ in Fig.~3 is split at the AFM ordering for spin-up
and spin-down states, and this provides the formation of the magnetic
moments on the iron atom. As is seen from Table~3, it is accompanied
by a noticeable decrease in $N(E_{\rm F})$ for the ferromagnetic and
antiferromagnetic phases of $R$Fe$_4$Al$_8$. The distinction between
the magnetic moments on a Fe atom for the FM and AFM phases appeared
insignificant in the case of LuFe$_4$Al$ _8$ and more noticeable in
ScFe$_4$Al$_8$. For all three investigated $R$Fe$_4$Al$_8$ compounds,
the minimum of the total energy is found for the AFM ordering of the Fe
moments in the basal plane along the [100] type directions. The
calculated values of magnetic moments from Table~3 are
qualitatively consistent with the experiments of Ref.
\cite{paixao01} for LuFe$_4$Al$_8$ ($\simeq 1.3 \mu_{\rm B}$) and
the results of calculations in \cite{gasche06} for YFe$_4$Al$_8$
($\simeq 1.25 \mu_{\rm B}$).

The calculated value of DOS at the Fermi level for the AFM phase of
LuFe$_4$Al$_8$, \mbox{$N(E_{\rm F})=14.3$} states/(eV$\cdot$\,cell),
can be compared with experimental data on the electronic specific
heat coefficient in this compound, $\gamma_{\rm
exp}=70$~mJ/mol\,$\cdot$\,K$^2$ \cite {dmitriev03}.\,\,According to
(\ref{gamma_lambda}), the corresponding renormalization parameter
for effective masses is about $\lambda \simeq 1$, which is
qualitatively consistent with the observation of superconductivity
in LuFe$_4$Al$_8$ \cite{dmitriev03}.\,\,It is necessary to consider,
however, a contribution of the spin-fluctuation term in (\ref{sf}),
which can be rather large for systems with high values of $N(E_{\rm
F}),$ to $\lambda$ \cite {grechnev09}.\,\,Therefore, the question
remains open: Can spin-fluctuations affect superconductivity in
$R$Fe$_4$Al$_8$ compounds and, if so, how?

\subsection{FeSe}
\begin{figure}%
\vskip1mm
\includegraphics[width=6cm]{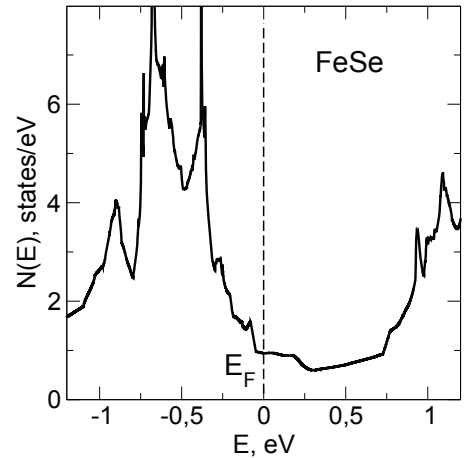}
\vskip-3mm\caption{Density of electronic states $N(E)$ of FeSe. The
Fermi level ($E=0$) is marked by the vertical line  }\label{fig4}
\end{figure}

The calculations of the electronic structure of FeSe were performed
for the tetragonal $P4/nmm$ structure and for the  orthorhombic
$Cmma$ structure, which correspond to the non-magnetic
superconducting phase. The crystal lattice parameters of FeSe were
taken according to data of Refs.
\cite{mizuguchi10,wen11,millican09,kumar10}. The calculated density
of electronic states of the tetragonal FeSe is presented in
Fig.~\ref{fig4}, where the dominating contribution to $N(E_{\rm F})$
comes from the \mbox{$3d$-states of iron.}

As is seen in Fig. \ref{fig4}, the Fermi level is located in a close
proximity ($\sim$0.1 eV) to the sharp peak of the density of
electronic states. A similar feature of $N(E)$ near $E_{\rm F}$
appears also in the orthorhombic phase of FeSe. It should be noted that
fine details of the electronic spectrum, in particular, positions of the
critical points of $E(k)$ relatively to $E_{\rm F}$, can be reliably
determined by the {\it ab initio} calculations within DFT with an
accuracy no more than 0.1~eV.

According to the results of calculations of the electronic structure
and the magnetic susceptibility in the normal state in an external
magnetic field, FeSe compound is very close to a magnetic
instability with the dominating exchange-enhanced spin paramagnetism
$\chi_{\rm spin}$. Within the Stoner model, the exchange-enhanced
Pauli spin contribution to the magnetic susceptibility can be
presented as $\chi_{\rm ston}=S\mu^2_{\rm B}N(E_{\rm F})$, where $S$
-- the Stoner factor, $N(E_{\rm F})$ -- the density of states at the
Fermi level, $\mu_{\rm B}$ -- the Bohr magneton.\,\,The calculated
value of DOS at the Fermi level for FeSe amounts to $N(E_{\rm
F})\simeq 1$ states/(eV$\cdot$cell), allowing a small uncertainty in
the determination of lattice parameters.\,\,Using the experimental
value of susceptibility of FeSe at low temperatures, $\chi \simeq
1.6\times 10^{-4}$~emu/mol \cite {grechnev13}, we obtained the
Stoner factor $S\simeq 5$.

The calculated $N(E_{\rm F})$ can be compared with experimental
data on the electronic specific heat coefficient in FeSe,
$\gamma_{\rm exp}=5.73$~mJ/mol$\cdot$K$^2$ \cite{lin11}. According
to (\ref{gamma_lambda}), the corresponding parameter of
renormalization of the effective masses amounts to $\lambda \simeq1.4$.
Assuming the validity of formalism (\ref{mcmillan})
for the superconducting transition in FeSe at 8~K and using the
experimental value $\Theta_{\rm D }=210$~K from \cite{lin11}, it is
possible to estimate the corresponding electron-phonon interaction
parameter, $\lambda_{\rm el-ph}\simeq0.9$. However, it is necessary
to consider also the spin-fluctuation term, according
to (\ref{sf}), which gives $\lambda_{\rm sf}\simeq0.5$.

The experimentally observed large effects of a pressure on the
superconducting transition temperature \cite{mizuguchi10,wen11} and
on the magnetic susceptibility \cite{grechnev13} gave evidence of
substantial changes in the electronic structure of FeSe under
pressure. In order to clarify a nature of these changes, the
geometry optimization was performed for the tetragonal $P4/nmm$
structure of FeSe within the GGA approach \cite{pbe96} and Elk
program \cite{elk}. In a such way, the pressure dependences of the
crystal lattice parameters were calculated. In particular, a growth
of the relative height $Z$ of selenium atoms over the iron atoms
plane under applied pressure was established. These results are in a
qualitative agreement with experimental data of \mbox{Refs.
\cite{millican09,kumar10}.}

\begin{figure}%
\vskip3mm
\includegraphics[width=6cm]{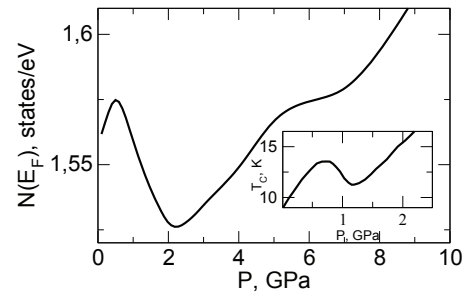}
\vskip-3mm\caption{Pressure dependence of the density of electronic
states of FeSe at the Fermi level, $N(E_{\rm F})$. The inset shows
the pressure dependence of the superconducting transition
temperature for FeSe taken from Ref. \cite{bendele12}  }\label{fig5}
\end{figure}

With the use of the established dependences of lattice parameters of
FeSe under pressure, we have calculated the corresponding behavior
of DOS at the Fermi level, $N(E_{\rm F },P)$, which is presented in
Fig. \ref{fig5}. It should be noted that the value of $N(E_{\rm
F},P=0)$ calculated within the GGA approach is by $\sim$1.5 times
larger than the corresponding value in Fig. \ref{fig4}, which was
calculated in the LDA approach \cite{barth72}. For the small
pressures ($0\div 0.3$ GPa, see Fig. \ref{fig5}), the calculated
derivative d$\ln N(E_{\rm F})/{\rm d}P\simeq 2.7\times 10^{-2}$
GPa$^{-1}$ is in a qualitative agreement with the experimental value
of the derivative of the magnetic susceptibility with respect to the
pressure at low temperatures, \mbox{d$\ln \chi/{\rm d}P\simeq
10$\,$\times$\,$10^{-2}$\,GPa$^{-1}$\,($0<p<0.2$\,GPa,
\cite{grechnev13})}, with regard for the Stoner factor of exchange
enhancement ($S\simeq 5$). Basically, the experimentally observed
large positive pressure effect on $\chi$ in FeSe at low temperatures
\cite{grechnev13} is caused by an increase of the internal
structural parameter $Z$ under pressure.

It should be noted that the calculated dependence of $N(E_{\rm F},P)$ in Fig. \ref{fig5}
correlates with the specific nonmonotonic behavior of the superconducting
transition temperature in the FeSe compound in a wide range of pressures \cite{bendele12}
(see inset in Fig. \ref{fig5}).
Such compliance can testify either in favor of the BCS mechanism of superconductivity,
or an alternative mechanism, which could involve, in a similar fashion,
the density of electronic states at the Fermi level.

\section{Conclusions}
The calculations of the densities of electronic states of $R$Ni$_2$B$_2$C, $R$Fe$_4$Al$_8,$ and FeSe
compounds indicate that the Fermi energy in these systems is located in a vicinity of
the pronounced peaks in $N(E)$.
This confirms the recent ARPES observations that the proximity of Van Hove spectral features
to the Fermi level can be considered as a key component for a realization of the superconductivity
in transition metal compounds \cite{kordyuk12}.
Though the main contribution to DOS for all three systems in a vicinity of $E_{\rm F}$ comes
from the quasi-two-dimensional layers of $3d$-metal atoms, nickel and iron, the $N(E)$ dependences
appear to be notably different.
The Fermi level in  $R$Fe$_4$Al$_8$ is located just at the peak of $N(E)$ and in the area of
high values of DOS. As for $R$Ni$_2$B$_2$C and FeSe compounds,
the Fermi level is actually in the ``pseudo-gap'' area of electronic spectra with
rather low values of $N(E_{\rm F})$.

The results of calculations of the electronic structure allow us to
analyze the experimental data on the electronic specific heat
coefficient (in $R$Ni$_2$B$_2$C, LuFe$_4$Al$_8$, FeSe) and the
cyclotron masses (DHVA effect in YNi$_2$B$_2$C and
LuNi$_2$B$_2$C).\,\,The estimated renormalization parameters of the
effective mass of conduction electrons indicate a possibility to
realize the electron-phonon mechanism of superconductivity in these
systems with $\lambda_{\rm el-ph}\simeq1$.\,\,Along with this, the
estimates indicate a noticeable contribution of the
electron-paramagnon (spin-fluctuation) interactions to $\lambda,$
which complies with a proximity of
 $R$Ni$_2$B$_2$C, $R$Fe$_4$Al$_8$ and FeSe to the magnetic or\-dering.

It is shown that the available experimental data on the strong pressure
dependence of the magnetic susceptibility and the superconducting
transition temperature in FeSe are caused by an increase in the density
of electronic states at the Fermi level under pressure. The
experimentally established nonmonotonic dependence of the
superconducting transition temperature on the pressure in FeSe qualitatively
correlates with the calculated behavior of DOS at the Fermi level in
a wide range of pressures (Fig.~5).

\vskip3mm {\it This work was supported by the Russian-Ukrainian
RFBR-NASU project 01-02-12.}

\rezume{%
Г.Є.\,Гречнєв, О.В.\,Логоша,\\ А.О.\,Легенька, О.Г.\,Гречнєв,
О.В.\,Федорченко}{ЕЛЕКТРОННА СТРУКТУРА I ВЛАСТИВОСТI\\ НОВIТНIХ
ШАРУВАТИХ НАДПРОВIДНИКIВ} {На основi теорiї функцiонала густини
(DFT) проведено систематичне вивчення електронної енергетичної
структури i магнiтних властивостей шаруватих надпровiдникiв
$R$Ni$_2$B$_2$C, $R$Fe$_4$Al$_8$ i FeSe. Обчислення дозволили
виявити ряд специфiчних особливостей електронної структури, що
можуть вiдповiдати за незвичайнi структурнi, магнiтнi та надпровiднi
властивостi цих систем. Продемонстровано, що енергiя Фермi
розташована в околi пiкiв густини електронних станiв (DOS). Головний
внесок в DOS на рiвнi Фермi дають $3d$-електрони. Обчислення ефектiв
тиску на електронну структуру i магнiтну сприйнятливiсть у
нормальному станi вказують, що в цих новiтнiх надпровiдниках домiнує
спiновий парамагнетизм, i вони близькi до магнiтної нестабiльностi.
Показано, що експериментальнi данi по залежностi температури
надпровiдного переходу в FeSe вiд тиску якiсно корелюють з
розрахованою поведiнкою DOS на рiвнi Фермi пiд тиском.}

\end{document}